\documentclass[10pt]{dis03}
\usepackage{amsmath}
\usepackage{graphicx}

\newcommand{\xgrv}{x_{\text{GRV}}}

\title{The triple-pole pomeron: Regge theory and DGLAP evolution}
\author{Gregory Soyez\footnote{G. Soyez is supported by the FNRS, Belgium; e-mail: {\em g.soyez@ulg.ac.be}}\\
{\small University of Liege, Institut de Physique, Bat B5, B4000 Liege, Belgium}}
\date{Talk presented for the DIS 2003 Workshop, 24-27 April 2003, St-Petersbourg, Russia}

\begin{document}

\maketitle

\begin{abstract}
We shall explain how it is possible to link Regge theory with DGLAP evolution using a triple-pole pomeron model. We shall first show that Regge theory can be used to constrain the initial condition for DGLAP evolution. We shall then spell out a method to extract Regge couplings at high $Q^2$ using DGLAP evolution.
\end{abstract}

\section{Introduction}

As it is well known, Regge theory and DGLAP evolution both provide very good descriptions of Deep Inelastic Scattering (DIS). One may therefore try to see if these two approaches are consistent\cite{Soyez:2002nm, Soyez:2003sr}. However, this question is far from evident due to the essential singularity generated by DGLAP evolution. 

Moreover, we have shown in \cite{Cudell:2002ej} that, as required by Regge theory, it is possible to describe all processes using the same singularity structure in complex $j$-plane. We shall show that this universal behaviour can be extended to parton distribution functions (PDF), i.e. that we can choose an initial condition for DGLAP evolution with a common singularity structure. In this picture, $F_2$ is describes by Regge theory for $Q^2<Q_0^2$ and by DGLAP equation for larger scales.

Finally, one has seen that, using a triple-pole pomeron, one can extend Regge theory to high values of $Q^2$, where we expect perturbative QCD (pQCD) to apply. In other words, the high-$Q^2$ behaviour of the Regge couplings must be predicted by DGLAP evolution. We shall see in section \ref{sec:resid} that it is the case for the triple-pole pomeron model.

\section{Constraints on initial condition}\label{sec:init}

\begin{figure}[ht]
\begin{center}
\includegraphics[scale=0.75]{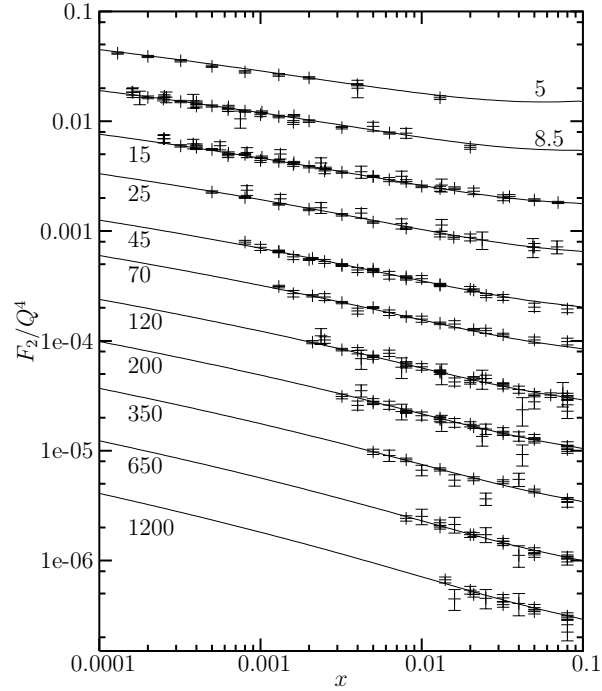}
\end{center}
\caption{Fit to $F_2^p$ structure function with $Q_0^2=5$ GeV$^2$ and $\xgrv=0.1$. $Q^2$ values (in GeV$^2$) are indicated for each curves.} \label{fig:init5}
\end{figure}

If we only want to reproduce $F_2^p$, we only need two quark distributions ($q^+=q+\bar{q}$): $T=x(u^++c^++t^+)-x(d^++s^++b^+)$ evolving alone and $\Sigma=x(u^++c^++t^+)-x(d^++s^++b^+)$ coupled with the gluon distribution $G=xg$. Using these definitions, the proton structure function becomes $F_2^p = \frac{5\Sigma+3T}{18}$.

If we introduce a scale $Q_0^2$ where DGLAP evolution breaks down, parton distributions will be described by a triple pole term and an $a_0/f$-reggeon term for $Q^2\le Q_0^2$ while an essential singularity arising from DGLAP evolution will be present at larger scales. Since Regge theory does not extend up to $x=1$, we shall use GRV parton distribution functions for $x>\xgrv$. Moreover, the pomeron contribution to $T$ must vanish because the pomeron is insentive to quark flavours and we shall assume that the reggeon, being mainly constituted of quarks, is not coupled to gluons. This leads to the following initial condition for DGLAP evolution ($\eta=0.31$)
\begin{eqnarray*}
     T & = & d_T x^\eta,\\
\Sigma & = & a_\Sigma \ln^2(1/x)+b_\Sigma \ln(1/x)+c_\Sigma+d_\Sigma x^\eta, \\
     G & = & a_G \ln^2(1/x)+b_G \ln(1/x)+c_G.
\end{eqnarray*}
In these expressions, the parameters in $\Sigma$ are constrained by the Regge fit of \cite{Cudell:2002ej} and, $d_T$ and $c_G$ are fixed by continuity with GRV at $x=\xgrv$. We are thus left with two parameters: $a_G$ and $b_G$.

We can adjust these parameters by fitting $F_2$ data from HERA, BCDMS, E665 and NMC within the domain\footnote{The first and second constraints come from Regge theory and the third one from DGLAP evolution.}
\begin{equation}\label{eq:domain}
\begin{cases}
\cos(\theta_t) =\frac{\sqrt{Q^2}}{2xm_p} \ge \frac{49\,\text{GeV}^2}{2m_p^2}\\
x\le\xgrv\\
Q_0^2\le Q^2\le 3000\,\text{GeV}^2
\end{cases}.
\end{equation}

One can show that for $Q_0^2\ge 3$ GeV$^2$ and $\xgrv\approx 0.1$, the fit gives a $\chi^2/nop$ smaller than 1.01. We have presented in Fig.\ref{fig:init5} the result for $Q_0^2=5$ GeV$^2$ and $\xgrv=0.1$, producing a $\chi^2/nop$ of 0.985.
Finally, one can show that the resulting gluon distribution is of the same order of magnitude as in the standard PDF sets.

\begin{figure}[ht]
\begin{center}
\includegraphics[scale=0.7]{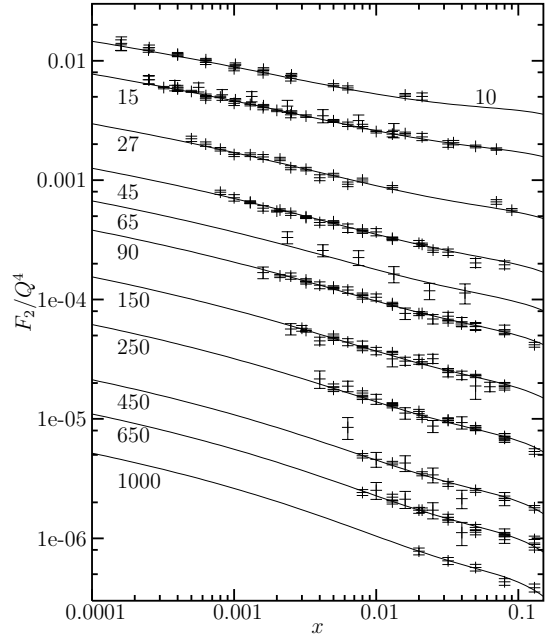}
\end{center}
\caption{Fit to $F_2^p$ structure function using the form factors extracted from our model and $\xgrv=0.15$. $Q^2$ values (in GeV$^2$) are indicated for each curve.} \label{fig:form5}
\end{figure}

\section{Regge couplings from DGLAP evolution}\label{sec:resid}

We know that the triple-pole pomeron model can be extended to high $Q^2$. In this case, the DGLAP evolution equation must be able to predict the residues at high $Q^2$. In such a case, due to the essential singularity generated by DGLAP evolution, we may only consider the DGLAP equation as a numerical approximation. In order to find the residues we may therefore adopt the following strategy:
\begin{enumerate}
\item\label{s21} choose an initial scale $Q_0^2$,
\item\label{s22} choose a value for the parameters in the initial distribution,
\item\label{s23} compute the parton distributions for $Q_0^2 \le Q^2  \le Q_{\text{max}}^2$ using forward DGLAP evolution and for $Q_{\text{min}}^2 \le Q^2  \le Q_0^2$ using backward DGLAP evolution,
\item\label{s24} repeat \ref{s22} and \ref{s23} until the value of the parameters reproducing the $F_2$ data for $Q^2>Q_{\text{min}}^2$ and $x\le \xgrv$ is found.
\item This gives the residues at the scale $Q_0^2$ and steps \ref{s21} to \ref{s24} are repeated in order to obtain the residues at all $Q^2$ values.
\end{enumerate}

\begin{figure}[ht]
\begin{center}
\includegraphics[scale=0.75]{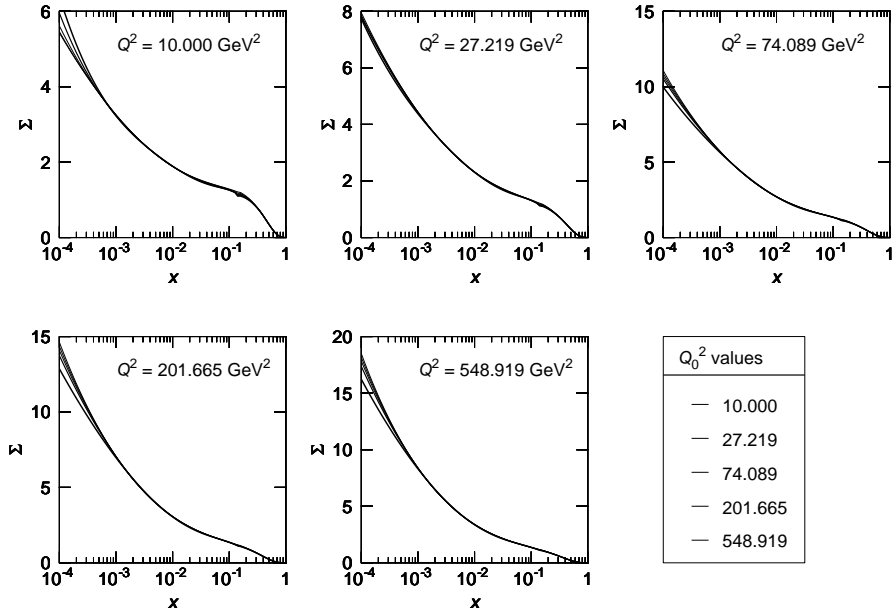}
\end{center}
\caption{Quark singlet distribution obtained from our model. Each curve corresponds to a choice of $Q_0^2$. $F_2$ is described by a triple pole if $Q^2=Q_0^2$ and by an essential singularity otherwise.} \label{fig:pdf}
\end{figure}

\begin{figure}[ht]
\begin{center}
\includegraphics[scale=0.75]{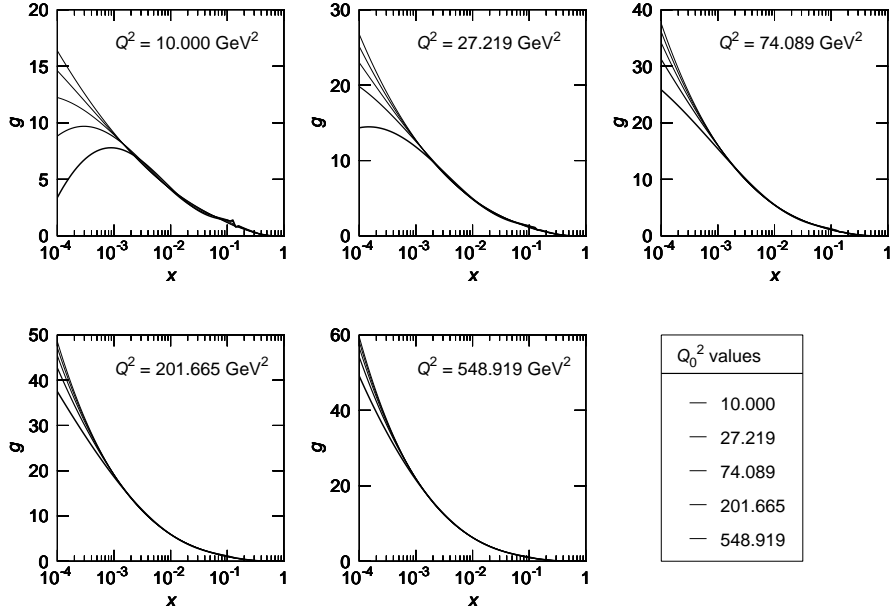}
\end{center}
\caption{Gluon distribution obtained with our method.} \label{fig:pdfg}
\end{figure}

As we can see in Fig. \ref{fig:form5}, the form factors found by this method reproduce $F_2$ successfully (For $10\le Q^2\le 1000$ GeV$^2$ and $\xgrv=0.15$, $\chi^2 = 576$ for 560 points).

Finally, we can look at the PDF obtained from this model. As shown in Fig. \ref{fig:pdf}, the singlet quark distribution nearly does not depend on $Q_0^2$, which proves that the DGLAP equation constitutes a good numerical approximation to the triple-pole pomeron. For the gluon distribution, we see from Fig. \ref{fig:pdfg} that things are different. However, all these gluon distributions give acceptable $F_2$ predictions and differences in the gluon distribution between different $Q_0^2$ must therefore be considered as uncertainties.

\section{Conclusion}

Firstly, we have seen that the Regge fit in \cite{Cudell:2002ej} can be successfully extended to high values of $Q^2$ using DGLAP evolution. In this method, quarks and gluons have a common singularity structure, as requested by Regge theory.

We have then shown that, assuming that Regge theory extends to high values of $Q^2$, one can consider DGLAP evolution as a numerical approximation and use it to extract the residues of the triple-pole pomeron at high $Q^2$. This method gives large uncertainties on the gluon distribution at low $x$ and low $Q^2$, which should be of prime importance for LHC physics.

In the future, we plan to see if we can extend this method to $x=1$ and to other Regge models. One should also try to find some modifications of the DGLAP equation stabilizing a triple-pole in $j=1$.

\end{document}